\documentclass[reprint,english,aps,prl,twocolumn,superscriptaddress]{revtex4-1}
\usepackage[T1]{fontenc}
\usepackage[latin9]{inputenc}
\usepackage{babel}
\usepackage{units}
\usepackage{textcomp}
\usepackage{amsmath}
\usepackage{amssymb}
\usepackage{graphicx}
\usepackage[unicode=true,pdfusetitle,
 bookmarks=true,bookmarksnumbered=false,bookmarksopen=false,
 breaklinks=false,pdfborder={0 0 1},backref=false,colorlinks=false]
 {hyperref}

\makeatletter

\usepackage{units}
\newcommand{\bra}[1]{\langle #1|}
\newcommand{\ket}[1]{|#1\rangle}

\usepackage{microtype}

\usepackage[T1]{fontenc}
\usepackage{lmodern}
\usepackage{mathbbol}

\makeatother

\begin{document}

\title{Symmetry-breaking supercollisions in Landau-quantized graphene}

\author{Florian Wendler}
\email{florian.wendler@fu-berlin.de}
\affiliation{Department of Theoretical Physics, Technical University Berlin, 10623 Berlin, Germany.}
\author{Martin Mittendorff}
\affiliation{Institute for Research in Electronics \& Applied Physics, University of Maryland, College Park, Maryland 20742, United States.}
\affiliation{Helmholtz-Zentrum Dresden-Rossendorf, PO Box 510119, D-01314 Dresden, Germany.}
\author{Jacob C. K\"onig-Otto}
\affiliation{Helmholtz-Zentrum Dresden-Rossendorf, PO Box 510119, D-01314 Dresden, Germany.}
\affiliation{Technische Universit\"at Dresden, D-01062 Dresden, Germany.}
\author{Samuel Brem}
\affiliation{Department of Theoretical Physics, Technical University Berlin, 10623 Berlin, Germany.}
\author{Claire Berger}
\affiliation{Georgia Institute of Technology, Atlanta, GA 30332, USA}
\affiliation{Institut N\'eel, CNRS-Universit\'e Alpes, 38042, Grenoble, France}
\author{Walter A. de Heer}
\affiliation{Georgia Institute of Technology, Atlanta, GA 30332, USA}
\author{Roman B\"ottger}
\affiliation{Helmholtz-Zentrum Dresden-Rossendorf, PO Box 510119, D-01314 Dresden, Germany.}
\author{Harald Schneider}
\affiliation{Helmholtz-Zentrum Dresden-Rossendorf, PO Box 510119, D-01314 Dresden, Germany.}
\author{Manfred Helm}
\affiliation{Helmholtz-Zentrum Dresden-Rossendorf, PO Box 510119, D-01314 Dresden, Germany.}
\affiliation{Technische Universit\"at Dresden, D-01062 Dresden, Germany.}
\author{Stephan Winnerl}
\affiliation{Helmholtz-Zentrum Dresden-Rossendorf, PO Box 510119, D-01314 Dresden, Germany.}
\author{Ermin Malic}
\affiliation{Department of Physics, Chalmers University of Technology, 41296 Gothenburg, Sweden.}

\begin{abstract}
Recent pump-probe experiments performed on graphene in a
perpendicular magnetic field have revealed carrier relaxation times ranging from
picoseconds to nanoseconds depending on the quality of the sample. To
explain this surprising behavior, we propose a novel symmetry-breaking
defect-assisted relaxation channel. This enables scattering of electrons with
single out-of-plane phonons, which drastically accelerate the carrier
scattering time in low-quality samples. The gained insights provide a strategy for
tuning the carrier relaxation time in graphene and related materials by
orders of magnitude.
\end{abstract}
\maketitle
In graphene, the impact of momentum-conserving acoustic phonon processes
on relaxation of non-equilibrium electronic excitations is limited by the low sound velocity, restricting
the maximum energy transfer per scattering event \cite{Levitov2012}.
Conversely, defect-assisted electron-phonon scattering events, so called supercollisions, can have a profound impact on the relaxation
time by removing the in-plane momentum conservation, since the excess momentum
can be absorbed by defects \cite{Levitov2012,Graham2013Nature,Betz2013,Graham2013Nano,Eless2013,Laitinen2014,Alencar2014}.
For graphene in a magnetic field, the situation is drastically changed since the in-plane translation symmetry is broken with the consequence
that the electron momentum is no longer a conserved quantity, suggesting that
supercollisions are not important in Landau-quantized graphene. 

In the light of this, the experimental results for electron relaxation times between graphene Landau levels in this letter
are very surprising, since they clearly show a strong dependence of the
relaxation time on the quality of the sample. To explain these findings,
we propose a novel mechanism for defect-assisted electron-phonon scattering breaking the mirror
symmetry of the graphene plane 
\footnote{In fact, breaking the mirror symmetry of the graphene plane means
that the momentum conservation in the out-of-plane direction is removed.}. 
By \textit{defect} we refer to local perturbations introducing additional
scattering channels and breaking the mirror symmetry. Such defects could e.g. be
local electric field gradients, interstitials, or bucklings of the
graphene plane induced by a substrate roughness or by vacancies \cite{Telling2003,Banhart2011}.
In general, the symmetry with respect to a mirror plane lying in the graphene plane prevents the linear coupling
of charge carriers and out-of-plane phonons (also known as flexural
phonons) in a perfect sample \cite{Mariani2008,Mariani2010_PHONONS}. The removal of the mirror symmetry by defects  
turns out to be crucial in Landau-quantized graphene, where carrier-phonon scattering is governed by resonances
between inter-Landau level transitions and corresponding phonon energies \cite{01_Wendler_Phonon_APL_2013,03_Wendler_Phonon_Pssb_2014}.
While the in-plane optical phonon energies of graphene range from
$150$ to \unit[200]{meV}, the out-of-plane phonon at the $\Gamma$
point (the $\Gamma\text{ZO}$ mode) -- activated by defect-electron interaction -- has an energy of \unit[100]{meV}
\cite{Maultzsch2004,Malic2011,Politano2015}. For a sufficiently large Landau level
(LL) broadening, supercollisions with $\Gamma\text{ZO}$ phonons allow transitions between the
three energetically lowest levels $\text{LL}_{\pm1}$ and $\text{LL}_0$ at reasonable magnetic fields.

\begin{figure}[!t]
\begin{centering}
\includegraphics[width=1\columnwidth]{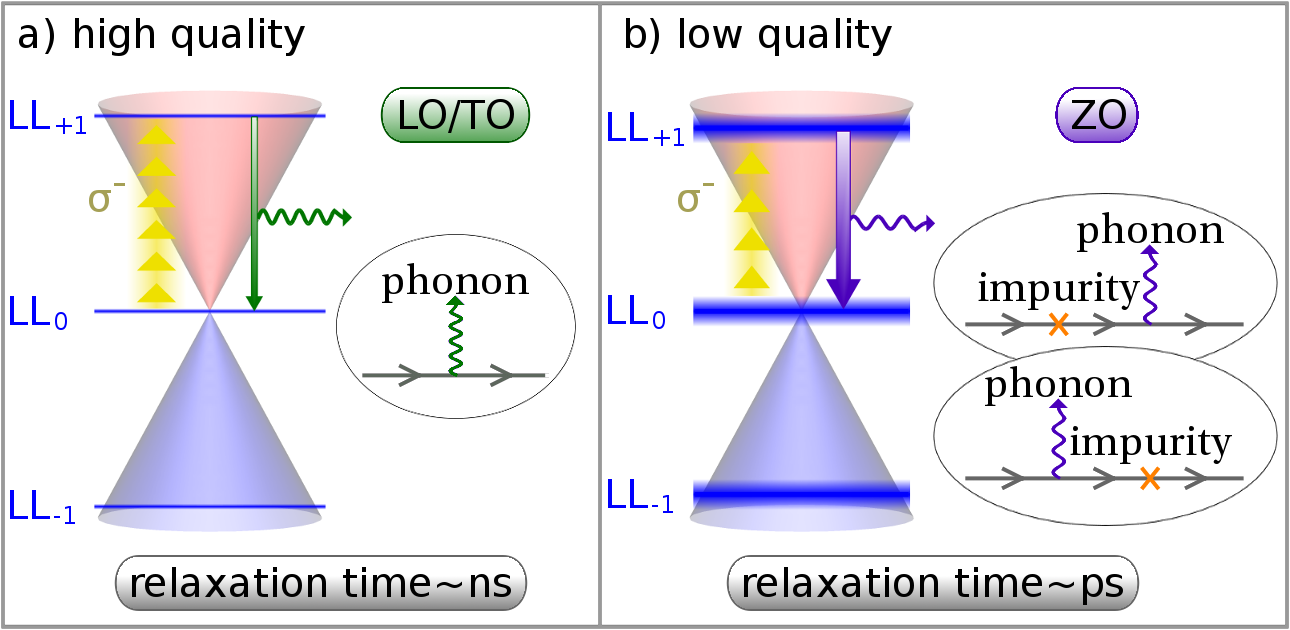}
\par\end{centering}

\protect\caption{Sketch of (a) carrier-phonon and (b) defect-assisted carrier-phonon
scattering channels governing the relaxation in high and low quality samples,
respectively. Three broadened Landau levels (LLs) are shown with the
Dirac cone in the background. Charge carriers are excited with left
circularly polarized radiation (yellow arrows) and relax via scattering
with phonons. Feynman diagrams of the scattering processes
illustrate the difference between (a) a high quality sample where usual carrier-phonon
scattering involving the modes $\Gamma\text{TO}$, KTO, $\Gamma\text{LO}$,
KLO takes place (green arrows), and (b) a low quality sample where the simultaneous
scattering of charge carriers with defects and phonons enable the
relaxation via out-of-plane phonons of the mode $\Gamma\text{ZO}$ (purple arrows).}

\label{fig:sketch}
\end{figure}
The sketches in Figure \ref{fig:sketch} compare the ordinary carrier-phonon
scattering with the defect-assisted carrier-phonon scattering in two samples with different qualities, where the quality is defined by the concentration of defects breaking the mirror symmetry of the graphene plane. The
investigated situation is very similar in both cases: The system is
excited using $\sigma^{-}$-polarized radiation that is in resonance
with the transition $\text{LL}_{0}\to\text{LL}_{+1}$, and it develops
back to equilibrium via the emission of phonons. While ordinary carrier-phonon
scattering (LO/TO) is possible in the high and in the low quality sample, the relaxation
in the low quality sample is governed by supercollisions allowing the emission
of out-of-plane phonons ($\Gamma\text{ZO}$). The latter phonons have a smaller energy that approximately fits the inter-LL distance of the experiments discussed below. Therefore the relaxation occurs on a picosecond timescale ($\tau\sim\unit[10]{ps}$) as opposed to a nanosecond timescale ($\tau\sim\unit[1]{ns}$) in case of a high quality sample. 

\begin{figure}[!t]
\begin{centering}
\includegraphics[width=1\columnwidth]{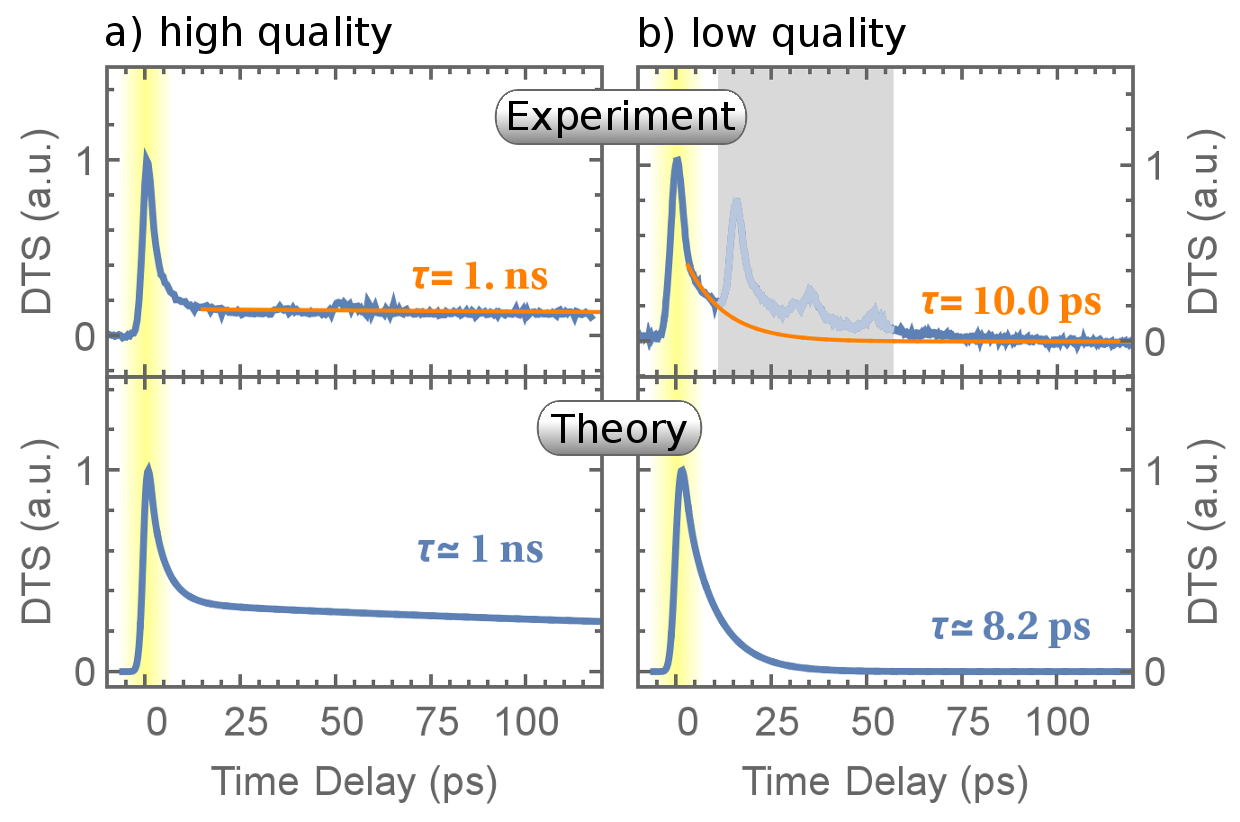}
\par\end{centering}

\protect\caption{Direct comparison of the carrier relaxation times seen in differential transmission spectra (DTS) of (a) a high quality
and (b) a low quality sample in experiment and
theory, respectively. This demonstrates that the relaxation time is two orders of magnitude faster
in the low quality sample. The yellow area
in the background illustrates the width of the pump pulse. The orange signatures show exponential functions fitting the experimental data, and the gray area in the experimental signal indicates secondary peaks appearing
due to reflections of the pump pulse which are omitted in our analysis. Experimental DTS for the high quality sample is shown for a longer time range up to $\unit[1]{ps}$ in the supplemental material.}
\label{fig:relaxation time}
\end{figure}
We present a theoretical study supported by experimental observations. Degenerate pump-probe experiments were carried out on two different multilayer epitaxial graphene samples. To this end, mid-infrared radiation from a free-electron laser (photon energy 75 meV, pulse duration 2.7 ps, pump fluence $\unit[0.1]{\mu J/cm^2}$) of $\sigma^-$-polarization was employed for both pumping and probing. The samples were kept at $\unit[10]{K}$ in a magnetic field of $\unit[4.2]{T}$ perpendicular to the graphene layers. At this field, both pump and probe beam were resonant with the $\text{LL}_{0}\to\text{LL}_{+1}$ transition in the samples. A mercury-cadmium-telluride detector cooled by liquid nitrogen was used for low-noise detection.
The two multilayer epitaxial graphene samples were grown by the same technique, namely thermal decomposition of the C-face of semi-insulating 4H-SiC by the confinement controlled sublimation (CCS) method \cite{deHeer2011}, however under slightly different growth conditions. One sample, in the following named ``low quality sample'' has a low structural quality compared to standard CCS grown epitaxial multilayer graphene. The other one, in the following named ``high quality sample'' features an extraordinary structural quality with no D-peak in the Raman spectrum \cite{deHeer2010}. Details on the sample growth, characterization by Raman spectroscopy, sample doping and the carrier dynamics in absence of a magnetic field can be found in the supplemental material.

The applied many-particle theory is based on an expansion of the graphene Bloch equations
\cite{MalicBuch,KochBuch,08_Wendler_Review_Nanophotonics_2015} by out-of-plane symmetry-breaking electron-defect interactions. It reproduces well the surprising experimental behavior, where decay times in two samples differ by more than two orders of magnitude, cf. Fig. \ref{fig:relaxation time}. For the theoretical modeling a pump fluence
of $\epsilon_{\text{pf}}=1\cdot10^{-2}\text{\ensuremath{\mu}J/cm\texttwosuperior}$, and an defect-assisted level broadening of $\Gamma_{\text{de}}=0.1\text{meV}$ ($\Gamma_{\text{de}}=10\text{meV}$) is used for the high quality sample (low quality sample). In accordance with Ref. \cite{Suess2016}, the doping is set to $\mu=\unit[10]{meV}$ in the high quality case, while  a value of $\mu=\unit[28]{meV}$ is used in the low quality case as indicated by Ref. \cite{05_Mittendorff_Auger_NatPhys_2014}. Before the relaxation processes dominate, i.e. during the excitation pulse, peaks in the transmission featuring very fast decay times occur. They result from the optical excitation as well as Auger scattering. Since those have already been discussed in Ref. \cite{05_Mittendorff_Auger_NatPhys_2014}, we focus on the slower relaxation in this work. While the high quality
sample exhibits a very slow relaxation time of about $\tau\sim\unit[1]{ns}$,
the relaxation is much faster in case of the low quality sample reaching times in the range of $\tau\sim\unit[10]{ps}$. 

The explanation for the strong dependence of the relaxation time on
the sample quality is two-fold: First, as
was already mentioned above, the carrier-phonon relaxation rate
in Landau-quantized graphene is determined by a resonance condition of phonon energy and
the energy of the inter-LL transition. If both energies coincide for
a specific inter-LL transition, carrier-phonon scattering is efficient
between the two involved LLs. Second, the exact resonance condition is weakened
by a finite broadening that is given by the dephasing of Landau transitions 
which also determines the LL broadening \cite{10_Funk_broadening_PRB_2015}.
Therefore, off-resonant scattering is possible and its efficiency
increases with the broadening of the Landau levels. Consequently,
the relaxation is faster in low quality samples where the Landau level broadening
is larger. The resonance condition is determined by the magnetic field, responsible for the formation of Landau-states. For the LL-transition in a range of $\unit[100]{meV}$, investigated here, only flexural phonons ($\Gamma\text{ZO}$) provide a fitting energy channel. However, the corresponding electron-phonon scattering is symmetry forbidden in pure graphene \cite{Mariani2008}. Here, we propose that this relaxation channel is activated in low quality samples of Landau-quantized
graphene by enabling carrier-flexural phonon ($\Gamma\text{ZO}$) scattering due
to defect-assisted electron-phonon scattering. Only if both factors (a sufficient LL-broadening and the activation of flexural phonons) are taken into account, 
the experimentally observed dependence of the relaxation time on the
quality of the sample can be explained.

We proceed to set forth the details of the theory of symmetry-breaking supercollisions
in Landau-quantized graphene to explain the microscopic background of the observed surprising sample-dependent relaxation behavior. The defect-assisted carrier-phonon
scattering is described by the Hamiltonian
\begin{equation}
H=H_{0}+H_{\text{el-de}}+H_{\text{el-ph}},\label{eq:Hamiltonian}
\end{equation}
consisting of a free energy part $H_{0}=\sum_{i}\epsilon_{i}a_{i}^{\dagger}a_{i}+\sum_{\mathbf{p},\mu}\epsilon_{\mu\mathbf{q}_{\text{ph}}}\left(b_{\mu\mathbf{q}_{\text{ph}}}^{\dagger}b_{\mu\mathbf{q}_{\text{ph}}}+1/2\right)$
and two perturbations stemming from disorder $H_{\text{el-de}}=\sum_{i,f}D_{i,f}a_{f}^{\dagger}a_{i}$
and the electron-phonon coupling $H_{\text{el-ph}}=\sum_{i,f}\sum_{\mu\mathbf{q}_{\text{ph}}}g_{i,f}^{\mu,\mathbf{q}_{\text{ph}}}a_{f}^{\dagger}a_{i}\left(b_{\mu,\mathbf{q}_{\text{ph}}}+b_{\mu,-\mathbf{q}_{\text{ph}}}^{\dagger}\right)$.
Here, and in the remainder of this article, the electronic states are specified by a compound index $i$
comprising the spin $s_{i}=\pm1$, valley $\xi_{i}=\pm1$, band $\lambda_{i}=\pm1$,
Landau level index $n_i=0,\,1,\,2,\,\ldots$, and the quantum number
$m_i=0,\,1,\,2,\,\ldots N_{B}-1$ giving rise to the large LL degeneracy
$N_{B}=AeB/(2\pi\hbar)$ that scales linearly with the area of
graphene $A$, the elementary charge $e$ and the magnetic field
strength $B$. Neglecting the Zeeman effect and the spin-orbit interaction,
both of which are small as compared to the level broadening in graphene \cite{Goerbig2011Review}, the
spin and valley degrees of freedom also contribute to the degeneracy,
and the low-energetic Landau level spectrum is given by $\epsilon_{i}=\lambda_i v_{\text{F}}\sqrt{2n_i\hbar eB}$,
with the Fermi velocity \cite{CastroNeto2009} $v_{\text{F}}\approx\ensuremath{\unit[1]{nm/fs}}$.
Furthermore, the optical phonon is characterized by its mode $\mu$, momentum
$\hbar\mathbf{q}_{\text{ph}}$ and energy $\epsilon_{\mu\mathbf{q}_{\text{ph}}}\approx\epsilon_{\mu}$,
and the matrix elements $D_{i,f}$ and $g_{i,f}^{\mu,\mathbf{q}_{\text{ph}}}$
determine the probabilities of the electronic transition $i\to f$
due to electron-defect and electron-phonon scattering, respectively. 

Starting from the Hamiltonian Eq. \eqref{eq:Hamiltonian}, the supercollision
matrix element is obtained by a multiple scattering expansion \cite{GreensFunctionsBook}
to the lowest necessary order in the LL-basis (Fig. \ref{fig:sketch}b, inset)
\begin{equation}
\widetilde{g}_{i,f}=\bra{f}\left[H_{\text{el-de}}\, G_{0}^{\dagger}(\epsilon_{f})\, H_{\text{el-ph}}+H_{\text{el-ph}}G_{0}(\epsilon_{i})\, H_{\text{el-de}}\right]\ket{i},
\end{equation}
where $\ket{i}$, $\ket{f}$ correspond to initial and final Landau level states.
Introducing a finite broadening $\Gamma$ (its origin is discussed below), the free electron Green's
function reads $G_{0}(E)=\left(E-H_{0}+i\Gamma\right)^{-1}$, which
yields the supercollision matrix element
\begin{align}
\widetilde{g}_{i,f}^{\mu,\mathbf{q}_{\text{ph}}}\Big|^{\text{em/ab}} & \simeq\sum_{v}\Bigg[g_{i,v}^{\mu,\mp\mathbf{q}{}_{\text{ph}}}\frac{1}{\epsilon_{i}-\epsilon_{v}\mp\epsilon_{\mu,\mathbf{q}{}_{\text{ph}}}+i\Gamma}D_{v,f}\nonumber \\
 & \hspace{1.35cm}+D_{i,v}\frac{1}{\epsilon_{i}-\epsilon_{v}+i\Gamma}g_{v,f}^{\mu,\mp\mathbf{q}{}_{\text{ph}}}\Bigg],\label{eq:supercollision matrix element}
\end{align}
for the emission ($-$) and absorption ($+$) of the phonon ($\mu,\mathbf{q}_{\text{ph}}$).
In the regime of well-separated Landau levels, we set $D_{i,f}\to D_{i,f}\delta_{\lambda_{i},\lambda_{f}}\delta_{n_{i},n_{f}}$,
so that the denominator of the second term in Eq. \eqref{eq:supercollision matrix element}
reduces to $i\Gamma$. In a self-consistent Born-Markov approximation, the parameter $\Gamma$ is identified by the
quantum mechanical dephasing $\Gamma_{i,f}$ which determines the
LL-broadening \cite{10_Funk_broadening_PRB_2015} and will be explained
below. The discussion of the pure LL-defect $D_{i,f}$ and the pure LL-phonon interaction $g_{i,f}$ can be found in the supplemental material.

Using the supercollision matrix element from Eq. \ref{eq:supercollision matrix element},
we can calculate the transition probability for the emission and absorption
of phonons using the generalized Fermi's golden rule \cite{FlensbergBruusBook} 
\begin{align}
W_{i,f} & =\frac{2\pi}{\hbar}\sum_{\mathbf{\mu,\mathbf{q}_{\text{ph}}}}\left|\widetilde{g}_{i,f}^{\mu,\mathbf{q}_{\text{ph}}}\right|^{2}\label{eq:Fermi's golden rule}\\
 & \hspace{0.5cm}\left[\left(n_{\mu}+1\right)L_{\Gamma_{if}}\left(\triangle E_{i,f,\mu}^{\text{em}}\right)+n_{\mu}L_{\Gamma_{if}}\left(\triangle E_{i,f,\mu}^{\text{ab}}\right)\right],\nonumber 
\end{align}
with the energy differences $\triangle E_{i,f,\mu}^{\text{em/ab}}=\epsilon_{i}-\epsilon_{f}\mp\epsilon_{\mu}$
for the emission ($-$) and absorption ($+$) of a phonon, the Bose
distribution $n_{\mu}$,
and the Lorentzian $L_{\Gamma}\left(\varDelta E\right)=\Gamma/\left(\pi\left(\varDelta E^{2}+\Gamma^{2}\right)\right)$
expressing energy conservation. Then, the carrier dynamics becomes
accessible via the graphene Bloch equations \cite{MalicBuch,08_Wendler_Review_Nanophotonics_2015}
\begin{align}
\dot{\rho}_{i} & =\sum_{j}\left[W_{j,i}\rho_{j}\left(1-\rho_{i}\right)-W_{i,j}\rho_{i}\left(1-\rho_{j}\right)\right],\label{eq:phonon part of d/dt rho}\\
\dot{p}_{i,f} & =-\frac{\Gamma_{i,f}}{\hbar} p_{i,f},\label{eq:phonon part of d/dt p_if}
\end{align}
for the occupation probability $\rho_{i}(t)=\langle a_{i}^{\dagger}a_{i}\rangle(t)$
and the microscopic polarization $p_{i,f}(t)=\langle a_{f}^{\dagger}a_{i}\rangle(t)$,
where Eq. \ref{eq:phonon part of d/dt rho} is equivalent to the Boltzmann
equation. The electron-light, electron-electron and ordinary electron-phonon interactions are
omitted for reasons of clarity. Their contribution to the carrier
dynamics is taken into account in the numerical evaluation, but it
has been explained elsewhere \cite{08_Wendler_Review_Nanophotonics_2015,05_Mittendorff_Auger_NatPhys_2014,02_Wendler_CM_NatCommun_2014}.
The total dephasing rate $\Gamma_{i,f}=\Gamma_{i,f}^{\text{de}}+\Gamma_{i,f}^{\text{el-el}}+\Gamma_{i,f}^{\text{el-ph}}+\Gamma_{i,f}^{\text{el-de-ph}}$
determines the broadening of the LLs and is composed of the defect-induced
dephasing $\Gamma_{i,f}^{\text{de}}$, and the dephasing contributions
due to the different many-particle scattering channels. For details
about $\Gamma_{i,f}^{\text{de}}$, $\Gamma_{i,f}^{\text{el-el}}$
and $\Gamma_{i,f}^{\text{el-ph}}$ we refer to Ref. \cite{10_Funk_broadening_PRB_2015}.
The dephasing induced by supercollisions is given by 
\begin{equation}
\Gamma_{i,f}^{\text{el-de-ph}}=\sum_{j}\left[\left(W_{j,i}+W_{j,f}\right)\rho_{j}+\left(W_{i,j}+W_{f,j}\right)\left(1-\rho_{j}\right)\right].
\end{equation}
Note that the scattering rates given by Eq. \ref{eq:Fermi's golden rule}
depend on the total dephasing $\Gamma_{i,f}$ which is needed to calculate
the dephasing due to supercollisions (and all the other many-particle
scattering contributions as well). Therefore, the total dephasing
is calculated self-consistently in every time step of the numerical
evaluation of the Bloch equations (Eqs. \ref{eq:phonon part of d/dt rho}-\ref{eq:phonon part of d/dt p_if})
\footnote{To this end, starting with an initial dephasing $\Gamma_{i,f}|^{0}$,
a sequence of dephasings $\Gamma_{i,f}|^{1}$, $\Gamma_{i,f}|^{2},\,\ldots$
is calculated where the preceding dephasing $\Gamma_{i,f}|^{l-1}$
is used to calculate the dephasing $\Gamma_{i,f}|^{l}$, respectively,
until two consecutive dephasings are equal.}.

\begin{figure}[!t]
\begin{centering}
\includegraphics[width=1\columnwidth]{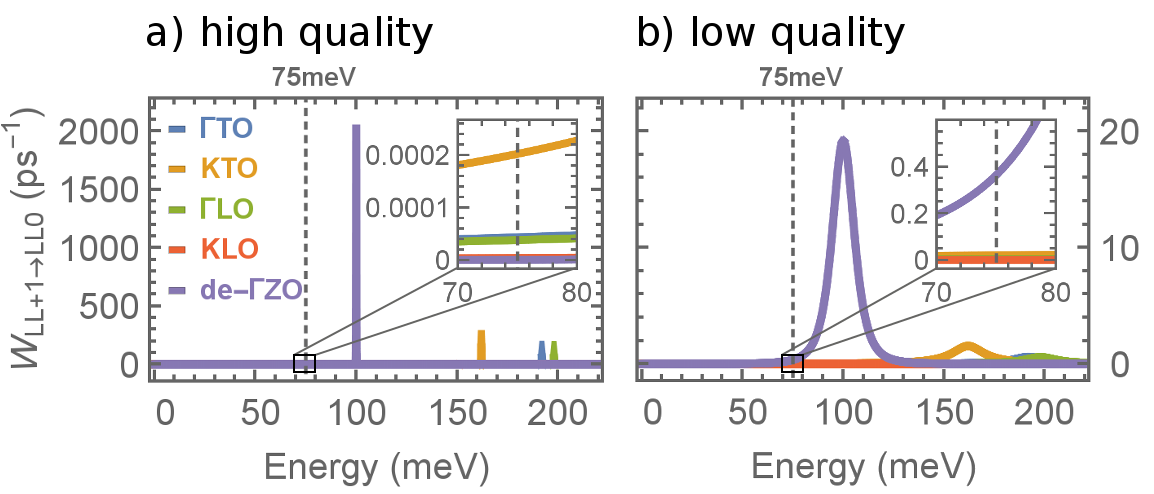}
\par\end{centering}

\protect\caption{Scattering rates induced by different phonon modes in dependence
of the energy of the inter-LL transition showing clear electron-phonon
resonance peaks for (a) the high quality ($\Gamma_{\text{de}}=0.1\text{meV}$)
and (b) the low quality sample ($\Gamma_{\text{de}}=10\text{meV}$). At a
magnetic field of $B=\unit[4.2]{T}$ (used in the experiments), the
relevant inter-LL spacings are $\unit[75]{meV}$ (dotted vertical
lines). The insets show that in this energy range defect-assisted electron-$\Gamma\text{ZO}$-phonon scattering ($\text{de-}\Gamma\text{ZO}$) clearly dominates in the low quality sample, while it is negligible in the high quality sample. 
}

\label{fig:scattering rates}
\end{figure}
Before we examine the numerical results in more detail, we investigate
how the scattering rates scale with the defect-induced 
dephasing which is proportional to the defect-induced LL broadening. In case of highly off-resonant electron-phonon scattering
($\Gamma_{i,f}\ll\varDelta E$), the absolute square of the supercollision
matrix element (Eq. \ref{eq:supercollision matrix element}) scales
like $|\widetilde{g}_{i,f}^{\mu,\mathbf{q}_{\text{ph}}}|^{2}\propto\Gamma_{\text{de}}^{2}/\Gamma_{i,f}^{2}$,
the Lorentzians in the scattering rate (Eq. \ref{eq:Fermi's golden rule})
can be approximated as $\pi L_{\Gamma_{i,f}}\left(\varDelta E\right)\simeq\Gamma_{i,f}/\varDelta E^{2}$,
and the scattering rate hence scales like $W_{i,f}\propto\Gamma_{\text{de}}^{2}/\Gamma_{i,j}$.
Moreover, in this off-resonant regime, the self-consistently determined
dephasing $\Gamma_{i,f}$ is generally of the same order of magnitude
as the defect-induced dephasing $\Gamma_{i,f}^{\text{de}}$, which
yields an effective linear scaling of $W_{i,f}$ with $\Gamma_{i,f}^{\text{de}}$.
Consequently, to approximate the scattering rate $W_{\text{LL}_{+1}\to\text{LL}_{0}}$
for supercollisions as well as the other relevant electron-phonon
scattering channels, we set $\Gamma_{i,f}\to\Gamma_{i,f}^{\text{de}}$.
The result is shown in Fig. \ref{fig:scattering rates} and demonstrates
that normal electron-phonon scattering is suppressed mainly due to
the high optical phonon energies, and likewise the smaller energy
of the $\Gamma\text{ZO}$-mode explains its importance for the relaxation
in Landau-quantized graphene. Note, however, that, as long as we are in the
regime of well-separated Landau levels, the energy of acoustic phonons
is too small to have an impact on the carrier dynamics, cf. Ref \cite{08_Wendler_Review_Nanophotonics_2015}.
Furthermore, Fig. \ref{fig:scattering rates} shows that, in the high quality sample, the supercollisions are strongly suppressed in the relevant (off-resonant) energy range, cf. inset of Fig. \ref{fig:scattering rates}a, since the off-resonance enters not only through the Lorentzian broadening in Eq. \ref{eq:Fermi's golden rule} (like in the case of ordinary electron-phonon scattering) but also through the energy differences in the supercollision matrix element (Eq. \ref{eq:supercollision matrix element}). In the low quality sample, on the other hand, the total dephasing is so large that the relevant energy lies within the broadened resonance peak of the supercollisions (purple line in Fig. \ref{fig:scattering rates}b), which is thus the dominant relaxation channel in this case. Therefore, the sample quality determining the total dephasing (via $\Gamma_{i,f}^{\text{de}}$), critically influences the relaxation.
This is the key to understand the
drastic difference in the relaxation times of the two experimentally
investigated graphene samples, cf. Fig. \ref{fig:relaxation time}.
Note that the scattering rates shown in Fig. \ref{fig:scattering rates} are calculated under
the assumption that the total dephasing $\Gamma_{i,f}$ equals the
defect-induced dephasing $\Gamma_{i,f}^{\text{de}}$, which is a
good approximation as long as we are in the off-resonant regime ($\Gamma_{i,f}\ll\varDelta E$).
The peak heights are thus not to be trusted and are shown only to
illustrate the energy detuning of the relevant inter-LL spacings and
the phonon modes.

\begin{figure}[!t]
\begin{centering}
\includegraphics[width=1\columnwidth]{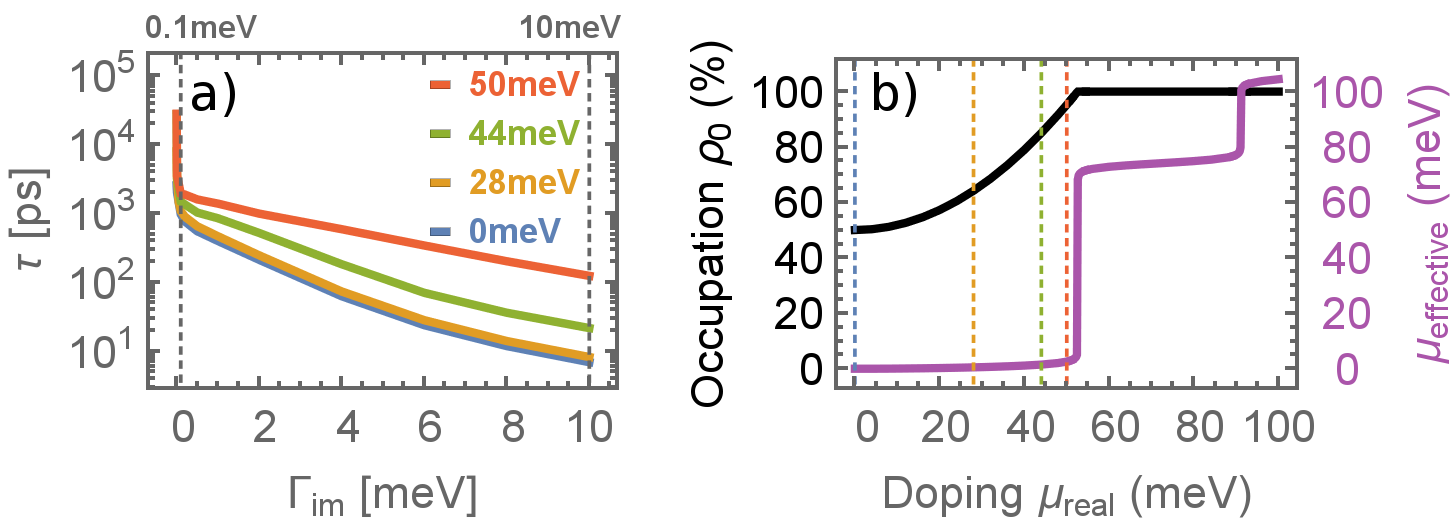}
\par\end{centering}

\protect\caption{(a) Relaxation time as a function of the defect-induced dephasing for different dopings $\mu$. The dashed vertical lines
mark the values of the defect-induced dephasing that were
used to model the high quality ($\Gamma_{\text{de}}=0.1\text{meV}$) and
the low quality ($\Gamma_{\text{de}}=10\text{meV}$) samples. (b) Initial
occupation probability of $\text{LL}_{0}$ in dependence of the doping
of the sample, and effective doping for a magnetic field of $B=\unit[4.2]{T}$
assuming that the carrier density does not change when the magnetic field
is turned on. The dashed vertical lines mark the values of the doping
that are used in (a).}

\label{fig:relaxation time (Gamma_im)}
\end{figure}

The dependence of the relaxation time on the defect-induced dephasing, which is a measure for the sample quality, is shown
in Fig. \ref{fig:relaxation time (Gamma_im)} (a) for different values
of the doping. It shows a very slow relaxation for samples with zero
disorder $\tau\sim\unit[10]{ns}$ which becomes about three orders of
magnitude faster as the disorder concentration is increased. 
The very strong decrease of $\tau$ for vanishing defect-induced dephasings $\Gamma_{\text{de}}<\unit[0.1]{meV}$ occurs due to the fact, that in this case the defect-induced dephasing is no longer the dominant contribution to the total dephasing $\Gamma$. Instead, in the regime  $\Gamma_{\text{de}}\ll\Gamma$, the total dephasing is essentially determined by electron-electron scattering that does not change the occupations -- known as pure-dephasing --, but constitutes a scattering channel which competes with relaxation channels (like electron-phonon scattering and supercollisions).
Furthermore,
we observe that the relaxation time depends on the doping of the sample,
since this determines the initial occupations and therewith the Pauli
blocking, cf. the occupations in Eq. \ref{eq:phonon part of d/dt rho}.
Considering solely the transition $\text{LL}_{+1}\to\text{LL}_{0}$,
Eq. \ref{eq:phonon part of d/dt rho} reads $\dot{\rho}_{+1}=-W_{+1\to0}\rho_{+1}\left(1-\rho_{0}\right)$,
which -- assuming a constant $\rho_{0}$ -- yields an exponential
decay $\rho_{+1}=\exp(-t/\tau)$ with the relaxation time $\tau=1/(W_{+1\to0}(1-\rho_{0}))$.
This illustrates that $\tau$ can become very large for a nearly completely
filled $\text{LL}_{0}$. To understand the minimal change of the relaxation
time for dopings below $\sim28\text{meV}$ and the marked change for
higher dopings, we take a look at the initial value of the occupation
$\rho_{0}$ in dependence of the doping of the sample, cf. black line
in Fig. \ref{fig:relaxation time (Gamma_im)} (b), resulting from
a Fermi distribution for an effective doping (purple line). The effective
doping $\mu_{\text{effective}}$ for a magnetic field of $B=\unit[4.2]{T}$
is calculated from the doping in the absence of a field $\mu_{\text{real}}$
using the criterion that the carrier density does not change when
the magnetic field is switched on, cf. Ref. \cite{05_Mittendorff_Auger_NatPhys_2014}.
The observation that $\text{LL}_{0}$ is far from being completely
filled for low dopings, while it is almost filled for dopings higher
than $\mu_\text{real}\sim40\text{meV}$, explains the doping dependence of the relaxation
time.

Finally, we compare the obtained results to magnetotransport experiments from Refs. \onlinecite{Baker2012, Baker2013}, where the temperature dependence of energy loss rates in a variety of Landau-quantized graphene samples was measured. 
The results are in agreement with ordinary supercollisions involving acoustic phonons, and the high-field energy loss rates are only about 40 \% smaller in comparison to the zero-field case. 
This seems to contradict our results but in fact it is consistent with our interpretation.
The samples used in the aforementioned magnetotransport experiments are single layer graphene samples with a finite doping, where the Fermi energy lies at the 8th Landau level or higher. 
Therefore, according to the findings of Ref. \onlinecite{Orlita2011}, the experiments are performed in a regime where the Landau levels are not well separated, i.e. adjacent Landau levels overlap.
Consequently, scattering with low energetic acoustic phonons and ordinary supercollisions have a profound impact on the relaxation dynamics, but supercollisions do no dramatically change the carrier dynamics. 
On the other hand, the present paper studies one particular inter-Landau level transition ($\text{LL}_{0}\to\text{LL}_{+1}$), where an energy gap of a few tens of meV efficiently supresses acoustic phonon scattering (too low energy) and also in-plane optical phonon scattering (too high energy).
Here, the activation of the $\Gamma\text{ZO}$ phonon mode having an intermediate energy can lead to a dramatically different relaxation dynamics.

In summary, the relaxation time in Landau-quantized graphene shows
a very distinct dependence on the sample quality. This is explained by a novel defect-assisted
electron-phonon scattering channel, in which disorder breaks the mirror symmetry
of graphene. These findings demonstrate that varying the amount of
disorder can be used to tailor the relaxation time in this system
over several orders of magnitude.

\begin{acknowledgments}
We acknowledge financial support from the Deutsche Forschungsgemeinschaft DFG for support through SFB
658 and SPP 1459, the EU Graphene Flagship (contract no. CNECT-ICT-604391), the Swedish Research Council (VR), and the NSF through grant \#1506006.
Support by the Ion Beam Center (IBC) at HZDR is gratefully acknowledged.
Furthermore, we thank A. Knorr (TU Berlin) for fruitful discussions on the carrier dynamics in graphene, and we are grateful to J. Maultzsch (TU Berlin) and E. Mariani (University of Exeter) for inspiring discussions on symmetry and flexural phonons.
\end{acknowledgments}

%

\end{document}


\title{Supplemental material for ``Symmetry-breaking supercollisions in Landau-quantized graphene''}

\author{Florian Wendler}
\email{florian.wendler@fu-berlin.de}
\affiliation{Department of Theoretical Physics, Technical University Berlin, 10623 Berlin, Germany.}
\author{Martin Mittendorff}
\affiliation{Institute for Research in Electronics \& Applied Physics, University of Maryland, College Park, Maryland 20742, United States.}
\affiliation{Helmholtz-Zentrum Dresden-Rossendorf, PO Box 510119, D-01314 Dresden, Germany.}
\author{Jacob C. K\"onig-Otto}
\affiliation{Helmholtz-Zentrum Dresden-Rossendorf, PO Box 510119, D-01314 Dresden, Germany.}
\affiliation{Technische Universit\"at Dresden, D-01062 Dresden, Germany.}
\author{Samuel Brem}
\affiliation{Department of Theoretical Physics, Technical University Berlin, 10623 Berlin, Germany.}
\author{Claire Berger}
\affiliation{Georgia Institute of Technology, Atlanta, GA 30332, USA}
\affiliation{Institut N\'eel, CNRS-Universit\'e Alpes, 38042, Grenoble, France}
\author{Walter A. de Heer}
\affiliation{Georgia Institute of Technology, Atlanta, GA 30332, USA}
\author{Roman B\"ottger}
\affiliation{Helmholtz-Zentrum Dresden-Rossendorf, PO Box 510119, D-01314 Dresden, Germany.}
\author{Harald Schneider}
\affiliation{Helmholtz-Zentrum Dresden-Rossendorf, PO Box 510119, D-01314 Dresden, Germany.}
\author{Manfred Helm}
\affiliation{Helmholtz-Zentrum Dresden-Rossendorf, PO Box 510119, D-01314 Dresden, Germany.}
\affiliation{Technische Universit\"at Dresden, D-01062 Dresden, Germany.}
\author{Stephan Winnerl}
\affiliation{Helmholtz-Zentrum Dresden-Rossendorf, PO Box 510119, D-01314 Dresden, Germany.}
\author{Ermin Malic}
\affiliation{Department of Physics, Chalmers University of Technology, 41296 Gothenburg, Sweden.}

\date{\today}

\pacs{}

\maketitle
\section*{Sample growth and doping}
Both multilayered epitaxial graphene (MEG) samples were grown on the C-terminated (000-1) face of 4H-SiC substrates with maximum 0.2 degree  misorientation by thermal decomposition of SiC. Two different confinement controlled sublimation (CCS) \cite{deHeer2011} furnaces were used, where graphene grows under the Si vapor in quasi-equilibrium above the SiC crystal at high temperature. The low quality sample was grown at $1440^\circ$C for 18 min on a $\text{H}_{2}$ pre-etched SiC substrate. The high quality sample was grown directly on an epitaxy ready CMP polished surface. It was annealed twice at $1400^\circ$C for 60 min, with a $\text{SiO}_{2}$ decomposition step at $1150^\circ$C for 20 min. 

The multilayer epitaxial graphene layers are composed of a highly doped layer at the interface with SiC, followed by about three layers of exponentially decreasing doping \cite{Sun2008} and large number (for our samples about 40 - 50 layers) of almost neutral graphene \cite{Orlita2008}. The experiments at a photon energy of 75\,meV are sensitive to the almost intrinsic layers. Experiments using all combinations of pumping and probing with $\sigma^+$ and $\sigma^-$ polarized radiation indicate that both samples feature slight n-type doping. The Fermi energy (at B = 0) for the high quality sample is about 10 \,meV to 15\,meV \cite{Koenig-Otto2016}, for the low quality sample $\sim$28\,meV \cite{Mittendorff2014}.

\section*{Long time scan}
\begin{figure}[b]
\includegraphics[trim=2.0cm 1.cm 0.cm 0.cm,width=0.4\textwidth]{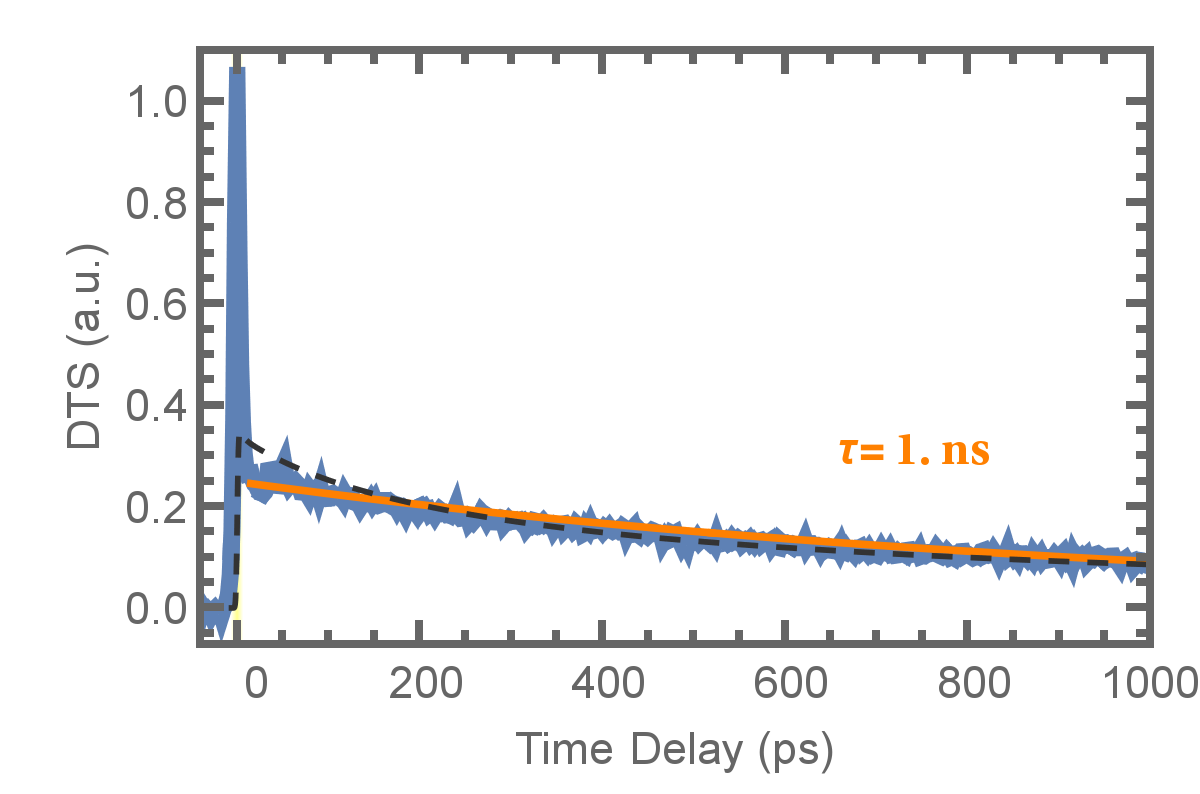}
	\caption{Differential transmission for linear polarized pumping and probing with $\sigma^{-}$-polarized radiation, in experiment (blue line), fit to experiment (orange line), and theory (dashed black line)}\label{fig:longTime}
\end{figure}

To demonstrate the slow decay time in the high quality sample, in Fig. \ref{fig:longTime} we provide the experimental signal for a time range of up to $\unit[1]{ns}$ (blue line) which is complemented by an exponential fit (orange line) and the theoretical result (dashed black line). Fig. \ref{fig:longTime} shows the result for linear polarized pumping and probing with $\sigma^{-}$-polarized radiation. The plot confirms the extracted decay time in the order of $\unit[1]{ns}$ in the high quality sample in Fig. \ref{fig:longTime}a in the main manuscript. Unfortunately, the data is not available for the most simple case discussed in the main manuscript, i.e. pumping and probing with $\sigma^{-}$-polarized radiation.

\section*{Raman spectroscopy}
\begin{figure}[t]
\includegraphics[trim=0.0cm 1cm 0.1cm 1.2cm,width=0.4\textwidth]{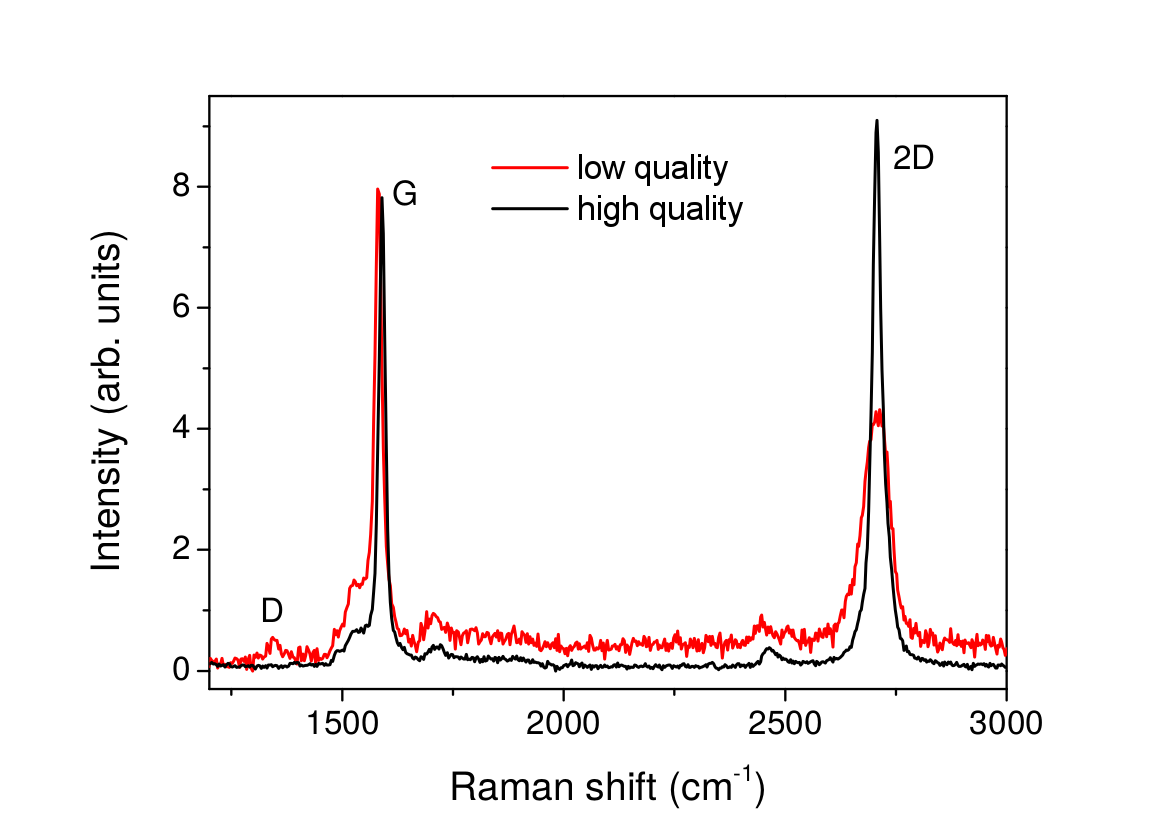}
	\caption{Raman spectra of the two samples.}\label{fig:Raman}
\end{figure}

The Raman spectra of both samples are depicted in Fig. \ref{fig:Raman}.  They are measured with an excitation wavelength of 532\,nm. Both samples feature single-Lorentzian 2D peaks as expected for electronically decoupled graphene layers \cite{Faugeras2008}. The most prominent indication for the high structural quality of the MEG samples is the low intensity of the D-peak at $1350\,\text{cm}^{-1}$, which is forbidden in defect-free graphene. Indeed the high quality sample shows no D-peak at all, showing the MEG excellent structural quality \cite{deHeer2010}. The spectrum of the low quality sample presents a very low D-peak, with an intensity ratio of the G-peak to the D-peak of 20. Note that the additional peaks observed around the G-peak arise from the SiC substrate, and are strongly attenuated because of the large number of graphene layers. A further indication of high structural quality is the narrow width of the 2D-peak. The full width at half maximum is $(24 \pm 2)\,\text{cm}^{-1}$ and $(56 \pm 11)\,\text{cm}^{-1}$ for the high quality and the low quality sample, respectively. The Raman analysis shows that the high quality sample has a structural quality that is comparable to the purest form of graphene \cite{Berciaud2009}, which is quite remarkable considering the large number of layers (The Raman spectrum probe them all since the substrate SiC peaks are visible). The low quality sample, on the other hand, shows signatures of a reduced structural quality, although the Raman spectrum still characterizes it as graphene with very low defect density. Graphene grown by chemical vapor deposition and transferred to $\text{SiO}_{2}$ is typically of much lower structural quality \cite{Lee2008}.

Note that the higher signal-to-noise ratio in the Raman spectrum of the high quality sample (cf. Fig. \ref{fig:Raman}) results from a longer integration time that was chosen to enable resolving even a very tiny D peak if it would be there. The pronounced difference of the G/2D ratios is mainly connected to differences in D-peak width. If we compare the ratios of peak areas, there is only a 5 \% difference in the ratio of the areas of the G peak and 2D peaks for the two samples.

\section*{Dynamics at zero magnetic field}
\begin{figure}[t]
\includegraphics[trim=0.0cm 1cm 0.1cm 1.2cm,width=0.4\textwidth]{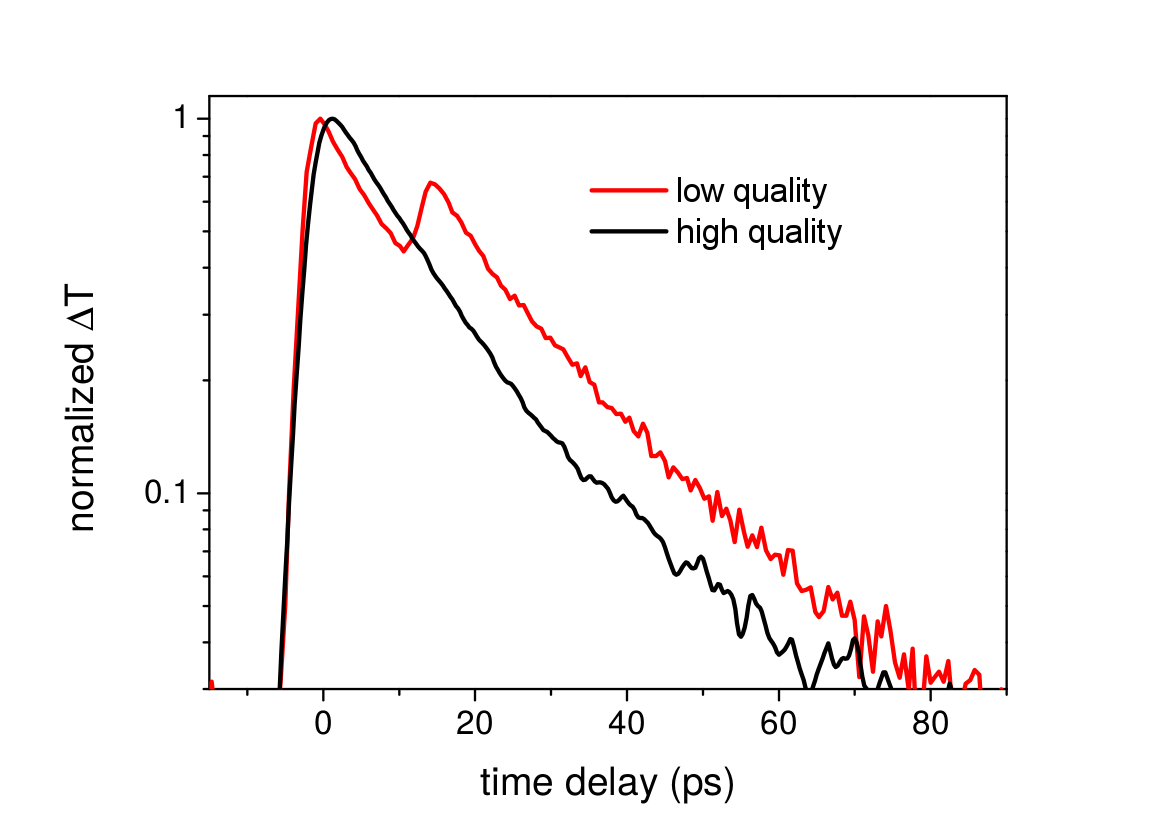}
	\caption{Normalized induced transmission for the two samples.}\label{fig:Pump_Probe}
\end{figure}

We investigated the carrier dynamics of both samples also in absence of a magnetic field. To this end, the same photon energy, pump fluence and substrate temperature as for the experiments under Landau quantization was applied. The experiments were performed with linearly polarized radiation with the polarization of the probe beam being perpendicular to the polarization of the pump beam. The results are depicted in Fig. \ref{fig:Pump_Probe}. Exactly the same relaxation time of $(13 \pm 2)$\,ps is found for the two samples. This is very surprising considering the fact that their relaxation times in case of Landau quantization differ by two orders of magnitude. This observation provides evidence for a very specific relaxation channel, which is opened up in the low quality sample in case of Landau quantization. 

Next, we discuss our results with respect to previous studies of the carrier dynamics in graphene with varied structural quality. In an early-stage degenerate pump-probe experiment at a photon energy of 1.5\,eV on multilayer epitaxial graphene on SiC, carrier cooling times between 0.4\,ps and 1.7\,ps for the lowest and highest quality sample, respectively, were found \cite{Dawlaty2008}. The much faster relaxation is due to efficient relaxation via optical phonons at the high photon energy \cite{Winnerl2011}. The observed dependence of the cooling time on the defect density may be explained by the fairly high amount of defects. The best sample in this study features a higher intensity ratio of G-peak to D-peak than our low quality sample. In a more recent study two-color pump-probe experiments have been performed on graphene with a systematically varied defect concentration \cite{Alencar2014}. In these experiment a pump photon energy of 3.1\,eV and a probe photon energy of 1.55\,eV was applied. Carrier cooling times ranging from 2.0\,ps for pristine graphene to 0.7\,ps for the sample with the highest amount of defects (intensity ratio of D to G-peak $\sim 1$) were measured. A large number of different samples was investigated in near-infrared pump – terahertz-probe experiments in a very recent study \cite{Mihnev2016}. This study is closest to our experiment for two reasons. Firstly, at the low probe photon energy relaxation via optical phonons is impossible. Consequently the observed timescales are much longer, in the 10\,ps range. Second, even though many different samples were investigated in the study, they were all of high structural quality as evidenced by a negligible D-peak in the Raman spectra, hence comparable to our two samples. Mihnev et al. do not observe differences in the relaxation dynamics for samples with slightly different structural quality. This behavior is corroborated by microscopic theory, which shows that defect related effects are negligible in this regime. The results of our experiments in absence of a magnetic field are in very good agreement with the findings of Mihnev et al \cite{Mihnev2016}.

\section*{Intentionally induced defects by ion irradiation}
In order to investigate the general role of defects on the carrier dynamics in Landau quantized graphene, a control experiment using ion irradiation was carried out. In this experiment some areas of the high quality sample, that had an area of $\unit[4\times5]{mm^2}$, were irradiated with C+ ions at an energy of $\unit[2]{keV}$. At this low energy, most of the ions are stopped within the 50 graphene layers and thereby intermixing at the interface to SiC is avoided. One area of the sample was irradiated with a dose of $\unit[1\times10^{12}]{cm^{-2}}$, another one with a dose of $\unit[5\times10^{13}]{cm^{-2}}$ and a third was kept pristine. Simple simulations, not taking into account healing effects, provide an upper limit for the concentration of generated defects. This is $\unit[6\times10^{11}]{cm^{-2}}$ for the lower dose. For the higher dose the concentration is 50 times higher. It is known that under these conditions mainly single and double vacancies are formed \cite{Lehtinen2010,Ahlgren2011}. Such point defects can enhance defect-assisted electron-phonon scattering, the supercollision process in absence of Landau quantization \cite{Alencar2014}. However, since they are not symmetry breaking, these defects should not result in symmetry-breaking supercollisions in Landau-quantized graphene.

\begin{figure}[t]
\includegraphics[width=0.5\textwidth]{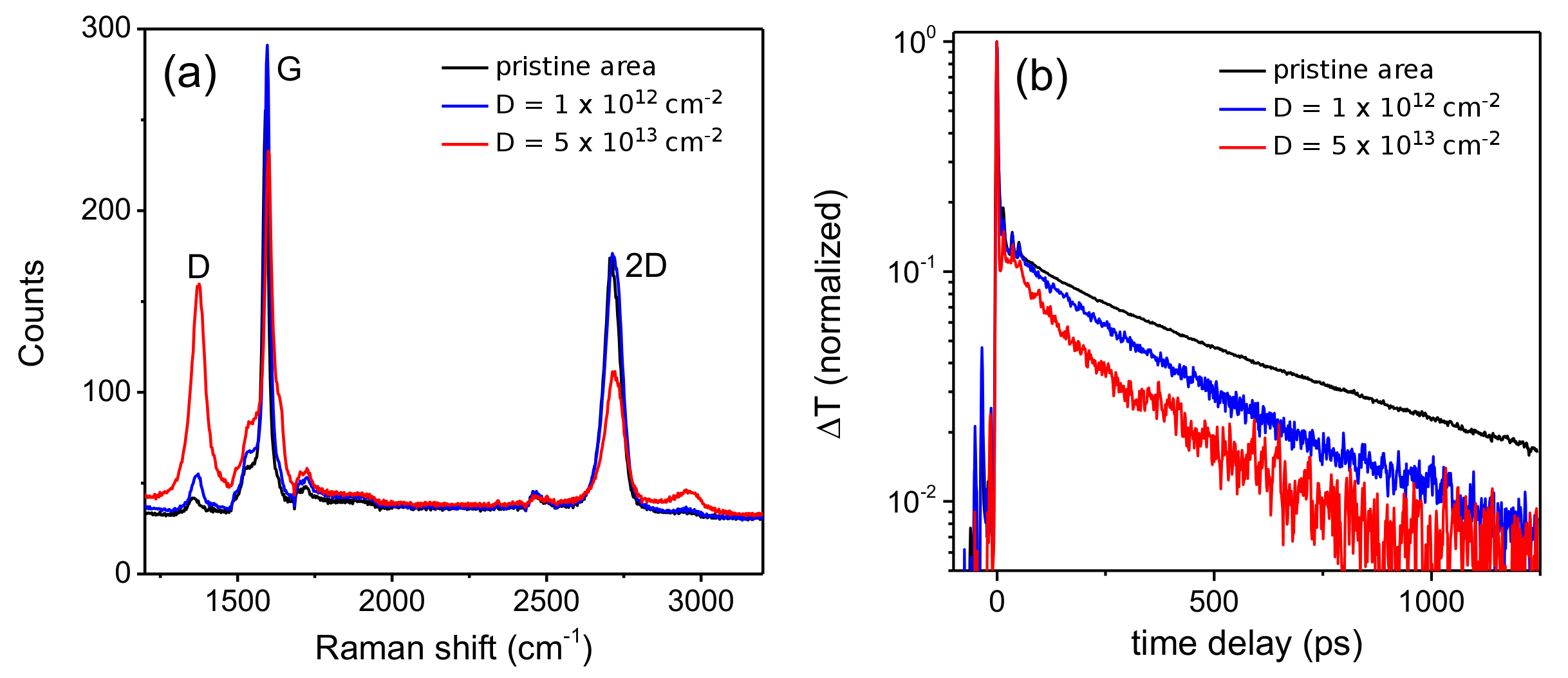}
	\caption{Raman spectra (a) and pump-probe signals (b) obtained at a magnetic field of $\unit[4.1]{T}$ for the ion-irradiated sample employed in the control experiment.}\label{fig:induced_defects}
\end{figure}

In Fig. \ref{fig:induced_defects}a Raman spectra from the three areas are depicted. The very small D peak for the pristine graphene, that is absent in the spectrum depicted in Fig. \ref{fig:Raman} seems to indicate a slight aging of the sample over the roughly 5 years that have passed between the measurements. The induced defects by ion irradiation are clearly seen in the Raman spectra with the dose of $\unit[1\times10^{12}]{cm^{-2}}$  ($\unit[5\times10^{13}]{cm^{-2}}$) resulting in a G/D peak ratio of 13 (1.6). Note that even for the low dose, the G/D peak ratio of 13 is lower than the ratio of 20 found for the “low quality sample” that is assumed to contain symmetry-breaking defects.

Degenerate pump-probe experiments applying $\sigma^-$ polarized pump and probe beams and a photon energy of $\unit[75]{meV}$ were carried out in the resonant magnetic field of $\unit[4.1]{T}$. Clearly the relaxation time is reduced with increasing amount of defects. However, the cooling times remain in the few $\unit[100]{ps}$ range (cf. Fig. \ref{fig:induced_defects}b). Even for the case of implantation with a dose of $\unit[5\times10^{13}]{cm^{-2}}$, the relaxation in the time window from $\unit[30-300]{ps}$ is about $\unit[200]{ps}$, i.e. 20 times larger than for the “low quality” sample described in the main text (cf. Fig. 2). Note that at the same time the G/D peak ratio of the sample implanted with high dose is $\sim13$ times lower than for the “low quality” sample. This is direct evidence for two facts: Firstly, there is a bottleneck in carrier cooling in Landau quantized graphene when the levels are spaced non-resonantly with the optical phonon energies. Relaxation times in the range for few $\unit[100]{ps}$ are found even in a sample with a very large concentration of point defects, so this bottleneck is very robust. Secondly, the results indicate that the “low quality sample” must feature an unusual mechanism that leads to the very short relaxation time of $\unit[10]{ps}$. Finding direct experimental evidence for the presence of symmetry breaking defects and understanding their formation during the growth process are important tasks, which go beyond our present study, in which we present mainly theory that provides a very strong indication that rapid relaxation is due to relaxation via out-of-plane phonons mediated by symmetry-breaking defects. Further experimental work is needed to unambiguously prove that symmetry-breaking supercollisions with ZO phonons are responsible for the dramatically different decay times.

\section*{Details of the theory}
The disorder averaged (denoted by $\overline{\cdots}$, cf. Ref. \onlinecite{Akkermans2007}) electron-defect matrix element reads 
\begin{align}
\overline{\left|D_{i,f}\right|^{2}} & =\frac{\Gamma_{\text{de}}^{2}}{N_{B}}\frac{\alpha_{n_{i}}^{2}\alpha_{n_{f}}^{2}}{4}\\
 & \hspace{0.5cm}\sum_{\mathbf{q}_{\text{de}}}\left[\left|F_{n_{f},m_{f}}^{n_{i},m_{i}}(\mathbf{q}_{\text{de}})\right|^{2}+\lambda_{i}^{2}\lambda_{f}^{2}\left|F_{n_{f}-1,m_{f}}^{n_{i}-1,m_{i}}(\mathbf{q}_{\text{de}})\right|^{2}\right],\nonumber 
\end{align}
with the definition $\alpha_{n}(n=0)=\sqrt{2}$, $\alpha_{n}(n\neq0)=1$,
the form factor $F_{nm}^{n'm'}(q,\varphi)$  which can be found in
Ref. \cite{08_Wendler_Review_Nanophotonics_2015}, and the defect-induced
dephasing $\Gamma_{\text{de}}$ \cite{10_Funk_broadening_PRB_2015}.
Note that in order to allow the symmetry-breaking supercollisions
proposed in this work, the disorder needs to break the mirror symmetry
of the graphene plane. Consequently, we only consider defects of
that kind, but in general it is conceivable that there are also defects
that affect the dephasing without contributing to supercollisions.
It would be straightforward to generalize our approach to this case.

Since the in-plane phonon modes were already described in detail in Ref. \cite{08_Wendler_Review_Nanophotonics_2015}, we focus here on the out-of-plane $\Gamma\text{ZO}$-mode.
The corresponding electron-phonon matrix element is given by \cite{Mariani2010_PHONONS,08_Wendler_Review_Nanophotonics_2015}
\begin{align}
\left|g_{i,f}^{\Gamma\text{ZO},\mathbf{q}_{\text{ph}}}\right|^{2} & =\frac{\hbar^{2}v_{\text{F}}^{2}\lambda_{\Gamma\text{ZO}}}{A}\frac{\alpha_{n_{i}}^{2}\alpha_{n_{f}}^{2}}{4}\label{eq:g_GZO squared under Landau-quantization}\\
 & \hspace{0.5cm}\left|F_{n_{f},m_{f}}^{n_{i},m_{i}}(\mathbf{q}_{\text{ph}})+\xi_{i}\xi_{f}\lambda_{i}\lambda_{f}F_{n_{f}-1,m_{f}}^{n_{i}-1,m_{i}}(\mathbf{q}_{\text{ph}})\right|^{2},\nonumber 
\end{align}
where $\lambda_{\mu,\mathbf{q}_{\text{ph}}}=F_{\mu,\mathbf{q}_{\text{ph}}}^{2}/(M\epsilon_{\mu,\mathbf{q}_{\text{ph}}}v_{\text{F}}^{2})$
is the dimensionless coupling strength as defined in Ref. \cite{Basko2009},
with the force exerted on electrons upon a deformation (corresponding
to the phonon $\mu,\mathbf{q}_{\text{ph}}$) $F_{\mu,\mathbf{q}_{\text{ph}}}$
and the mass density of graphene $M=\unit[7.6\times10^{-8}]{gcm^{-2}}$
\cite{Malic2011}. The calculation of the coupling strength of an
electron with a single $\Gamma\text{ZO}$-phonon ($\lambda_{\Gamma\text{ZO}}$)
is not straightforward, since this scattering process is forbidden
in free-standing graphene and becomes allowed, only if the symmetry
with respect to out-of-plane deformations is broken \cite{Mariani2008}.
Provided this mirror symmetry is broken, for
example due to the presence of a substrate or due to an external electric
field, a deformation in one perpendicular direction with respect to the graphene plane gives a different energy
contribution to electrons than the same deformation in the reverse direction.
This means that a deformation in one out-of plane direction $\text{d}z$
leads to a force $\left(F_{x},F_{y},F_{z}\right)$ being exerted on
the electrons in graphene, which does not simply equalize the mirrored
force, i.e. $\left(F_{x},F_{y},-F_{z}\right)$, under the reverse
deformation $-\text{d}z$. Therefore, the coupling strength $\lambda_{\Gamma\text{ZO}}$
is determined by that force and it depends on the concrete origin
of the mechanism breaking the mirror symmetry. We estimate the coupling
strength to be $\lambda_{\Gamma\text{ZO}}=0.0435$, a value recently
measured for graphene on Pt(111) by Politano et al. \cite{Politano2015}
in the first experiment demonstrating a first-order
coupling of electrons and out-of-plane-phonons in graphene %
\footnote{In fact, the paper states that $\lambda_{\Gamma\text{ZO}}=0.087$.
However, within our notation it should be $\lambda_{\Gamma\text{ZO}}=0.087/2$.%
}. While this is certainly a rough estimate, the exact value of the
coupling strength is not vital for the novel supercollision
we are proposing, since the strong impact of the out-of-plane phonons
mainly relies on their favorable energy.

Note that out-of-plane acoustic (ZA) phonons have a very low energy at the $\Gamma$ point rendering them inefficient for the decay between the lowest Landau levels (where the Landau level distance is the largest). ZA phonons at the K point, have a larger energy and they could result in a faster decay in our theory, which would indicate that the mirror-breaking defect concentration is lower than assumed (but still of the same order). However, we want to stress that the message of the paper is unaffected by this phonon mode.

\section*{Phonon spectrum in multilayer graphene}
As mentioned above, it has been shown that the individual graphene layers are essentially decoupled in terms of their electronic band structure due to small in-plane rotations of the different layers \cite{Faugeras2008}. However, as Ref. \onlinecite{Cocemasov2013} suggests, this is not necessarily true for the phonon spectrum. The paper theoretically investigates the phonon dispersion for twisted graphene layers, where the increased unit cell size leads to the emergence of a large number of hybrid folded phonon modes. This provides an alternative explanation for the dramatically different decay times in the two samples we investigated: A different twisting angle leads to different phonon energies, so that a hybrid folded phonon mode in one of the samples could be in resonance with the inter-Landau level transition under investigation, while this is not the case in the other sample. However, both multilayered epitaxial graphene samples were grown on the C-terminated (000-1) face of 4H-SiC substrates. For these samples it is known that the different graphene sheets can have the rotation angles $30\degree$ ($R30$) or $30\pm2.204\degree$ ($R2^\pm$), where $R30$ and $R2^\pm$ graphene sheets are interleaved to produce a high density of $R30/R2^\pm$ fault pairs instead of occurring as AB stacked $R30$ or $R2^\pm$ isolated domains \cite{Hass2008}. Therefore, the structure of both samples with respect to the rotation angles between the graphene sheets is expected to be the same, hence, the explanation of our experimental results via different rotation angles in the two samples is unlikely to be true.

Moreover, our results do not rely on the exact phonon energies, but on the fact that the optical in-plane phonon energies are too high and the acoustic in-plane phonon energies are too low to explain the fast decay in the low quality sample, while the $\Gamma\text{ZO}$-mode has an intermediate energy in the range of the relevant inter-Landau level transition. From the observed extremely slow decay in the high quality sample we can conclude that phonon modes with energies fitting the inter-Landau level transition under investigation are absent in this sample, hence, they should also be absent in the low quality sample.

\FloatBarrier
%